# Using Space-Time trellis Codes For AF Relay Channels

Sue Keramat, Zue Xie

November 19, 2018

**Abstract**

We consider the analysis and design of space-time trellis codes (STTCs) for a cooperative relay channel operating in amplify-and-forward (AF) mode assuming the source and destination nodes are equipped with multiple antennas but the relay node has single antenna. We derive a pairwise error probability (PEP) expression for the performance of STTCs in this type of channels. A simple upper-bound on PEP is then derived and is maximized to find the optimum STTCs. We show that the designed STTCs based on the derived criterion achieve full diversity in the AF relay channels especially at high signal-to-noise-ratios (SNRs). The maximum achievable diversity in relay channels with single-antenna relay is bounded by $\min(M, N)$ where $M$ and $N$ are respectively the number of antennas in source and destination nodes. Simulation results confirm that the proposed codes achieve the maximum diversity and also provide an appealing coding gain.

## 0.1 Introduction

Cooperative communication is an effective way to improve diversity gain in wireless channels. Among cooperative communication schemes, the use of relaying has received a lot of attention in recent years [1]-[3], [11]. Most current research appears to focus on half-duplex operating relays as in the amplify-and-forward (AF) protocol because of its simplicity and feasibility. In the AF relaying protocol, often simply called AF protocol, the relay amplifies the received version of the transmitted source signal and transmits this amplified version to the destination.

On the other hand, multi-antenna techniques offer significant improvements in link reliability through the use of multiple antennas at the transmitter and/or receiver without incurring system losses in terms of delay and bandwidth efficiency. A significant advantage of multi-input-multi-output (MIMO) communications is the ability to provide a sound trade-off between diversity and multiplexing gains. The idea of using multiple transmit and receive antennas in wireless communication systems has attracted considerable attention with the aim of increasing data transmission rate and system capacity. A key issue is how to develop proper transmission techniques to efficiently exploit all the diversity gain offered by MIMO channels[3],[5],[6]. Space-Time Trellis Coding (STTC)[4] is an efficient transmission strategy in MIMO channels that enhances data rate and/or the reliability of communications over fading channels. Data is encoded by a trellis code and the encoded data is split into several streams that are then simultaneously transmitted using multiple transmit antennas. While the decoding complexity of STTCs is generally higher than linear Space-Time Block Codes (STBCs), they can provide larger diversity and coding gains [10].

While the design and evaluation of STTCs for point-to-point MIMO channels has been well investigated, the problem of STTC design for communication networks is still under study. There are several papers discussing space-time coding for cooperative communication [12]; but to the best of our knowledge, a general analysis on the design of STTCs in the cooperative relay systems operating in AF mode has not been done.

In this paper, the design and evaluation of STTCs for Rayleigh-faded cooperative channels with AF relaying is considered. As in many applications, it is desirable to keep the complexity at the relay node low, we consider single-antenna relay node throughout this paper. To design efficient STTCs for this scenario, pairwise error probability (PEP) is adopted as the performance measure. A proper metric for the design of STTCs in relay channels with multi-antenna source and destination and single-antenna relay nodes is derived based on an upper-bound on PEP. The derived bound is then used to design STTCs that outperforms the codes suggested in [4] in terms of bit-error-rate (BER). Finally, various numerical examples are provided to corroborate the analytical studies.

The remainder of this paper is organized as follows: a general model for AF relay channels with multi-antenna source and destination and single-antenna relay nodes is presented in Section II. In Section III, the PER criterion is in-troduced and an upper-bound on PEP is derived. The code design rule based on the derived upper-bound is then discussed. Simulation results for a variety of source and destination antenna combinations is presented in Section IV and Section V concludes the paper.

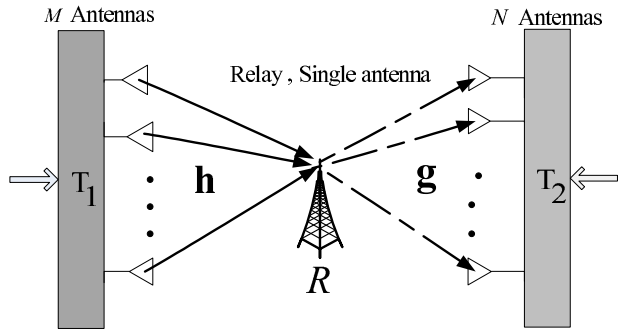

Figure 1: A MIMO cooperative scheme operating in amplify-and-forward (AF) mode with single-antenna relay node

## 0.2 System Model

Fig. 1 shows the general structure of the AF relay system under study. Node $T_1$ sends its information to node $T_2$ through an intermediate single-antenna relay node, R. It is assumed that R receives the signal from $T_1$ in one time-slot, amplifies the received signal and forwards it to $T_2$ in the next time-slot. We further assume that $T_1$ and $T_2$ are equipped with $M$ and $N$ antennas, respectively, and R is a single-antenna relay node. All the channels are assumed to be flat-fading. Under these conditions, the received baseband signal at R in the first time-slot is:

$$x = \mathbf{h}\mathbf{s} + n_1 \qquad (1)$$

where $x$ is the received signal at the relay; $\mathbf{s}$ is the $(M \times 1)$ transmitted vector from $T_1$; $n_1 \sim N(0, \sigma_1^2)$ is the complex circularly symmetric additive white Gaussian noise (AWGN) samples; and $\mathbf{h}$ is $(1 \times M)$ channel vector from $T_1$ to R. Upon receiving the signal from $T_1$, the relay R processes it with AF relay operation, and then broadcasts the processed signal to $T_2$ during the second time-slot. Mathematically, the linear processing operation at the relay can be concisely represented as $r = \alpha x$. For the sake of simplicity and without loss of generality, we assume $\alpha = 1$. The received signal in $T_2$ can then be written as:

$$\mathbf{y} = \mathbf{g}r + \mathbf{n}_2 \qquad (2)$$

where $\mathbf{n_2} \sim N(0, \sigma_3 \mathbf{I}_N)$ is the zero-mean circularly symmetric complex AWGN vector at node $T_2$. By substituting (1) in (2), the received signal at $T_2$ will be:

$$\mathbf{y} = \underbrace{\mathbf{ghs}}_{\text{signal}} + \underbrace{\mathbf{g}n_1 + \mathbf{n_2}}_{\text{noise}} \quad (3)$$

we can rewrite the above equation as following:

$$\tilde{\mathbf{s}} = \mathbf{Hs} + \hat{\mathbf{n}} \quad (4)$$

where $\mathbf{H} = \mathbf{gh}$ is a rank one matrix and

$$\hat{\mathbf{n}} \sim N(0, ((\sigma_1^2 \sigma_g^2 + \sigma_3^2)\mathbf{I}_N)) \quad (5)$$

where $\mathbf{H} = \mathbf{g}_r \mathbf{h}$ and $\hat{\mathbf{n}}_1 \sim N(0, (\sigma^2 \sigma_{g_r}^2 + \sigma_3^2)\mathbf{I}_N)$ where $\sigma_{g_r}^2 = E\{\mathbf{g_r g_r}^H\}$.

## 0.3 Pairwise Error Probability

PEP is the probability of a particular codeword $\mathbf{c}$ in a codeword set being decoded as another valid codeword $\mathbf{c}'$ at the receiver and is defined as [10]:

$$P(\mathbf{c} \to \mathbf{c}') = Q(\sqrt{\frac{E_s}{2N_0}} \|\mathbf{H}(\mathbf{c} - \mathbf{c}')\|^2) \quad (6)$$

where $E_s$ is the received signal energy, $N_0$ is the power spectral density of noise, and $Q(x)$ is the Gaussian Q-function defined as:

$$Q(x) \triangleq \frac{1}{\sqrt{2\pi}} \int_x^\infty e^{-\frac{\lambda^2}{2}} d\lambda \quad (7)$$

The probability of detection error for a particular codeword set $C$ can be written as the weighted sum of the PEP terms as:

$$P_e \leq \sum_{\mathbf{c} \in \mathbf{C}} P(\mathbf{c}) \sum_{\mathbf{c}' \neq \mathbf{c}} P(\mathbf{c} \to \mathbf{c}') \quad (8)$$

where $P(\mathbf{c} \to \mathbf{c}')$ represents the PEP and $P(\mathbf{c})$ is the probability of transmission of codeword $\mathbf{c}$. The PEP can be derived by using the Moment Generating Function (MGF) method [8]. The Q-function can be alternatively expressed in the Craig's formula as [7]:

$$Q(x) = \frac{1}{\pi} \int_0^{\frac{\pi}{2}} e^{-(\frac{x^2}{2\sin^2(\lambda)})} d\lambda \quad (9)$$

Integrating over all possible channel realizations results in an average PEP expression involving the MGF, i.e.,

$$P(\mathbf{c} \to \mathbf{c}') = \frac{1}{\pi} \int_0^{\frac{\pi}{2}} \mathrm{M}_{\|\mathbf{H}\Omega\|^2}\left(-\frac{E_s}{4N_0 \sin^2 \varphi}\right) d\varphi \leq \mathrm{M}_{\|\mathbf{H}\Omega\|^2}\left(-\frac{E_s}{4N_0}\right) \quad (10)$$

where $\mathrm{M}_{||\mathbf{H}.\mathbf{\Omega}||^2}(s) = E\{e^{||\mathbf{H}.\mathbf{\Omega}||^2 s}\}$ is the MGF of the random variable $||\mathbf{H}.\mathbf{\Omega}||^2$, $\mathbf{\Omega} = (\mathbf{c} - \mathbf{c}')$, and $\varphi$ is a dummy variable. The upper-bound $\mathrm{M}_{||\mathbf{H}\mathbf{\Omega}||^2}\left(-\frac{E_s}{4N_0}\right)$ for the average PEP in (10) is obtained using the Chernoff bound. Our objective is to find STTCs that minimize the upper-bound in (10) and hence will result in minimum average PEP. Note that this upper-bound includes the MGF of $||\mathbf{H}.\mathbf{\Omega}||^2$. Therefore, in the next step we analytically derive a closed-form expression for this MGF.

### 0.3.1 Calculation of Moment Generating Function

Using $\mathbf{H} = \mathbf{g}_r \mathbf{h}$, one can write:

$$\|\mathbf{g}_r \mathbf{h} \mathbf{\Omega}\|^2 = tr\left(\mathbf{g}_r \mathbf{h} \mathbf{\Omega} \mathbf{\Omega}^\mathbf{H} \mathbf{h}^H \mathbf{g}_r^H\right) = \|\mathbf{g}_r\|^2 \|\mathbf{h}\mathbf{\Omega}\|^2 = a.b \qquad (11)$$

Since $\mathbf{g}_r$ is a Gaussian vector with jointly independent elements, $a = \|\mathbf{g}_r\|^2$ has a chi-square distribution with $N$ degrees of freedom, i.e. $a \sim \chi^2_{2N}$. Let $\mathbf{\Omega}\mathbf{\Omega}^\mathbf{H} = \mathbf{U}^\mathbf{H} \mathbf{\Lambda} \mathbf{U}$ be the eigenvalue decomposition of $\mathbf{\Omega}\mathbf{\Omega}^\mathbf{H}$ in which $\mathbf{U}$ is a unitary matrix and $\mathbf{\Lambda} = diag(\lambda_1, ..., \lambda_M)$ where $\lambda_1, ..., \lambda_M$ are the eigenvalues of $\mathbf{\Omega}\mathbf{\Omega}^\mathbf{H}$. Since $\mathbf{\Omega}\mathbf{\Omega}^\mathbf{H}$ is symmetric positive definite, $\lambda_i$'s are non-negative. Also, the vector $\mathbf{g}$ is circularly symmetric Gaussian vector whose elements are jointly independent. We define:

$$b = \|\mathbf{h}\mathbf{\Omega}\|^2 = \sum_{i=1}^{M} \lambda_i |h_i|^2 \qquad (12)$$

In (12) the random variable $|h_i|^2$ is exponentially distributed; thus, $b$ has a non-central chi-square distribution. The MGF of $b$ can then be calculated as:

$$M_b(s) = E\{e^{sb}\} = \prod_{i=1}^{M} \frac{1}{1 - \lambda_i s} \qquad (13)$$

To derive the MGF of $\|\mathbf{H}(\mathbf{c} - \mathbf{c}')\|^2 = a.b$ we will use the following theorem.

**Theorem 1** *The MGF of $||\mathbf{H}.\mathbf{\Omega}||^2$ can be obtained as:*

$$M_{||\mathbf{H}.\mathbf{\Omega}||^2}(s) = \frac{1}{\Gamma(N)} \int_0^\infty \frac{x^{N-1}.e^{-x}}{\prod_{i=1}^{M}(1 - \lambda_i s x)} dx \qquad (14)$$

*where $\Gamma(N)$ is the standard Gamma function.*

**Proof 1** *Since $||\mathbf{H}.\mathbf{\Omega}||^2 = a.b$ where $a$ and $b$ are two independent positive random variables with probability density functions $f_a(.)$ and $f_b(.)$, it follows that*

$$M_{||\mathbf{H}.\mathbf{\Omega}||^2}(s) = \int_0^\infty e^{sv} f_{||\mathbf{H}.\mathbf{\Omega}||^2}(v) dv$$

$$= \int_0^\infty e^{sv} \int_0^\infty f_a(r) f_b(\frac{v}{r}) \frac{dr}{r} dv \stackrel{\frac{v}{r}=l}{=} \int_0^\infty f_a(r) \int_0^\infty e^{srl} f_b(l) dl \ dr \qquad (15)$$

As $a \sim \chi_{2N}^2$, $f_a(x) = \frac{x^{N-1}.e^{-x}}{\Gamma(N)}$. The second integral is the MGF of random variable b. Replacing $f_a(x)$ and the MGF of b, $M_b(sx)$, from (13) into (15), we obtain (14).

The integral in (14) cannot be calculated directly. Instead, we will use an approximation method in Section III.B to calculate it.

### 0.3.2 Approximation of Moment Generating Function

To analyze the system at high signal-to-noise-ratios (SNRs), we can look at the $M_{||\mathbf{H}.\mathbf{\Omega}||^2}(s)$ obtained in (14) for large values of $s$. We can then approximate the integral in (14) to find simple closed-form code design criteria. To this aim we state the following theorem.

**Theorem 2** *If all eigenvalues of $\mathbf{\Omega\Omega^H}$ are strictly positive and distinct; then $M_{||\mathbf{H}.\mathbf{\Omega}||^2}(s)$ can be approximated at large s as:*

$$M_{||\mathbf{H}.\mathbf{\Omega}||^2}(s) = \begin{cases} \frac{\Gamma(N-M)}{\Gamma(N) \prod_{i=1}^{M} \lambda_i} \frac{(-1)^M}{s^M} & N > M \\ \frac{1}{\Gamma(N) \prod_{i=1}^{M} \lambda_i} \frac{(-1)^M \log(-s)}{s^M} & N = M \\ (\frac{-1}{\Gamma(N)} \sum_{i=1}^{M} \frac{\log \lambda_i}{\lambda_i^N} (\prod_{i_1=1, i_1 \neq i}^{M} \frac{\lambda_i}{\lambda_i - \lambda_{i_1}})) \frac{1}{s^N} & N < M \end{cases} \qquad (16)$$

**Proof 2** *We consider two different cases: (a) $N > M$; and (b) $N \leq M$;*
*(a) For $N > M$, (14) can be rewritten as:*

$$M_{||\mathbf{H}.\mathbf{\Omega}||^2}(s) = \frac{1}{\Gamma(N) \prod_{i=1}^{M}(\lambda_i s)} \int_0^\infty \frac{x^{N-1}.e^{-x}}{\prod_{i=1}^{M}(\frac{1}{\lambda_i s} - x)} dx$$

$$\approx \frac{(-1)^M}{\Gamma(N) \prod_{i=1}^{M}(\lambda_i s)} \int_0^\infty x^{N-M-1}.e^{-x} dx$$

$$= \frac{(-1)^M}{\Gamma(N) \prod_{i=1}^{M}(\lambda_i s)} \Gamma(N-M) \qquad (17)$$

To derive the second equality, we use the fact that $\prod_{i=1}^{M}(\frac{1}{\lambda_i s} - x) \geq (-1)^M x^M$ for large s.

(b) For $N \leq M$, using the following identity ([9],chapter 7):

$$\int_0^\infty \frac{x^{N-1} \cdot e^{-x}}{\prod_{i=1}^M (1-\lambda_i sx)} \, dx =$$
$$\int_0^\infty \sum_{i=1}^M \left(\frac{x^{N-1}e^{-x}}{1-\lambda_i sx}\right) \left(\prod_{i_1=1, i_1 \neq i}^M \frac{\lambda_i}{\lambda_i - \lambda_{i_1}}\right) dx \quad (18)$$

(14) can then be rewritten as:

$$M_{\|\mathbf{H}.\mathbf{\Omega}\|^2}(s) =$$
$$\frac{1}{\Gamma(N)} \sum_{i=1}^M \left\{ \left(\prod_{i_1=1, i_1 \neq i}^M \frac{\lambda_i}{\lambda_i - \lambda_{i_1}}\right) \left(\int_0^\infty \frac{x^{N-1}e^{-x}}{1-\lambda_i sx} dx\right) \right\} \quad (19)$$

The integral in (19) behaves asymptotically as [9]:

$$\int_0^\infty \frac{x^{N-1}e^{-x}}{1-\lambda_i sx} \, dx \doteq$$
$$-\left(\sum_{j=1}^{N-1} \frac{\Gamma(N-j)}{\lambda_i^j s^j} + \frac{(\log s - \Gamma'(1))}{\lambda_i^j s^N} + \frac{\log \lambda_i}{\lambda_i^N s^N}\right) \quad (20)$$

where $\Gamma'(1)$ is differentiation of normal gamma function $\Gamma(x)$ in $x = 1$. By substituting the approximation (20) into (19), we have

$$M_{\|\mathbf{H}.\mathbf{\Omega}\|^2}(s) \doteq \sum_{i=1}^{N-1} \left(\sum_{j=1}^M \frac{1}{\lambda_j^i} \left(\prod_{i_1=1, i_1 \neq j}^M \frac{\lambda_j}{\lambda_j - \lambda_{i_1}}\right)\right) \frac{-\Gamma(N-i)}{s^i \Gamma(N)}$$
$$+ \frac{-(\log s - \Gamma'(1))}{\Gamma(N) s^N} \sum_{i=1}^{N-1} \left(\sum_{j=1}^M \frac{1}{\lambda_j^i} \left(\prod_{i_1=1, i_1 \neq j}^M \frac{\lambda_j}{\lambda_j - \lambda_{i_1}}\right)\right) \quad (21)$$
$$+ \frac{-1}{\Gamma(N) s^N} \left(\sum_{j=1}^M \frac{\log \lambda_j}{\lambda_j^N} \left(\prod_{i_1=1, i_1 \neq j}^M \frac{\lambda_j}{\lambda_j - \lambda_{i_1}}\right)\right)$$

If $N = M$, the first term becomes zero and for summation of two other terms we can write:

$$M_{\|\mathbf{H}\mathbf{\Omega}\|^2}(-s) \doteq \frac{(-1)^N}{\Gamma(N) \prod_{i=1}^N \lambda_i} \frac{(\log s - \Gamma'(1))}{s^N}$$
$$+ \frac{-1}{\Gamma(N)} \left(\sum_{j=1}^N \frac{\log \lambda_j}{\lambda_j^N} \left(\prod_{i_1=1, i_1 \neq j}^M \frac{\lambda_j}{\lambda_j - \lambda_{i_1}}\right)\right) \frac{1}{s^N} \quad (22)$$
$$\cong \frac{(-1)^N}{\Gamma(N) \prod_{i=1}^N \lambda_i} \frac{\log(-s)}{s^N}$$

If $N < M$, the second term in (21) is zero and the summation will be:

$$M_{\|\mathbf{H\Omega}\|^2}(s) \doteq$$
$$\frac{-1}{\Gamma(N)} \left( \sum_{j=1}^{M} \frac{\log \lambda_j}{\lambda_j^N} \left( \prod_{i_1=1, i_1 \neq j}^{M} \frac{\lambda_j}{\lambda_j - \lambda_{i_1}} \right) \right) \frac{1}{s^N} \qquad (23)$$

According to *Theorem 2*, the diversity order offered by a AF relaying with single-antenna relay is bounded by $\min(M, N)$. This is expected because this structure can be interpreted as the cascade of a multi-input-single-output (MISO) and a single-input-multi-output (SIMO) channel with maximum diversity order of $M$ and $N$, respectively. Besides, *Theorem 2* provides the maximum coding gain offered by this relay scheme which can be used in the design of better space-time codes for AF relaying with single-antenna relay.

### 0.3.3 Code Design

To design optimum STTCs for AF relaying with single-antenna relay, we consider two different cases: 1) For $N \geq M$, according to *Theorem 2*, the determinant criterion in [4] provides the optimum STTCs. 2) For $N < M$, the coding gain in (16) has a new form different from [4], and to minimize the PEP, the following metric should be minimized:

$$(-1)^{N-1} \sum_{i=1}^{M} \frac{\log \lambda_i}{\lambda_i^N} \left( \prod_{\substack{i_1=1 \\ i_1 \neq i}}^{M} \frac{\lambda_i}{\lambda_i - \lambda_{i_1}} \right) \qquad (24)$$

## 0.4 Conclusion

In this paper, the design of STTCs for AF relay channels with multi-antenna source and destination and single-antenna relay nodes was discussed. Based on the derived upper-bound for PEP, a code design rule was established for constructing optimal STTCs. The derived four-state QPSK codes with rate 2 bps/Hz for different numbers of transmit and receive antennas were shown to outperform the STTCs in [4] over Rayleigh fading channels. The performance gain obtained by using the proposed codes increases with the number of transmit antennas. Also, we proved that the diversity order offered by STTCs over an AF with single-antenna relay is bounded by $\min(M, N)$.

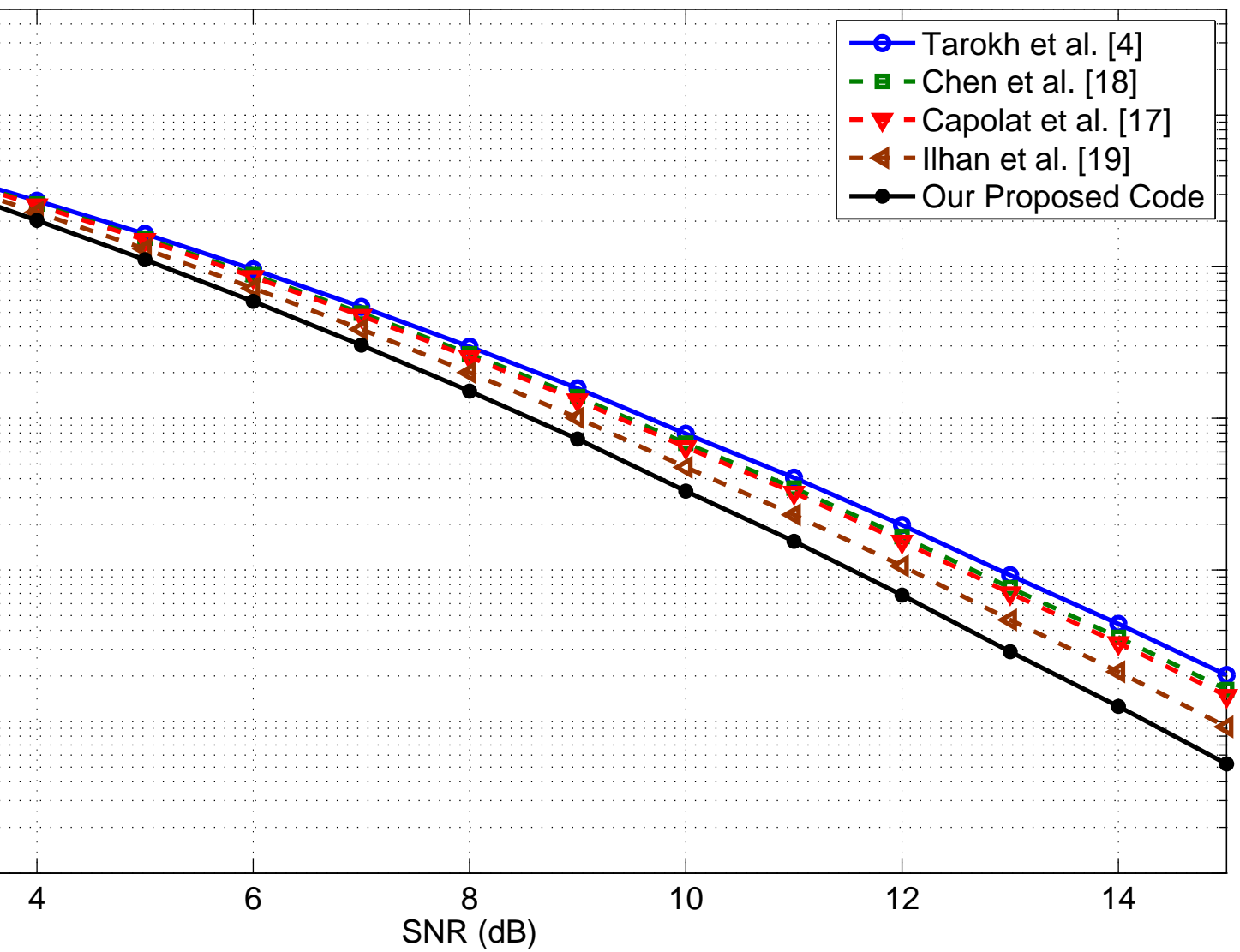

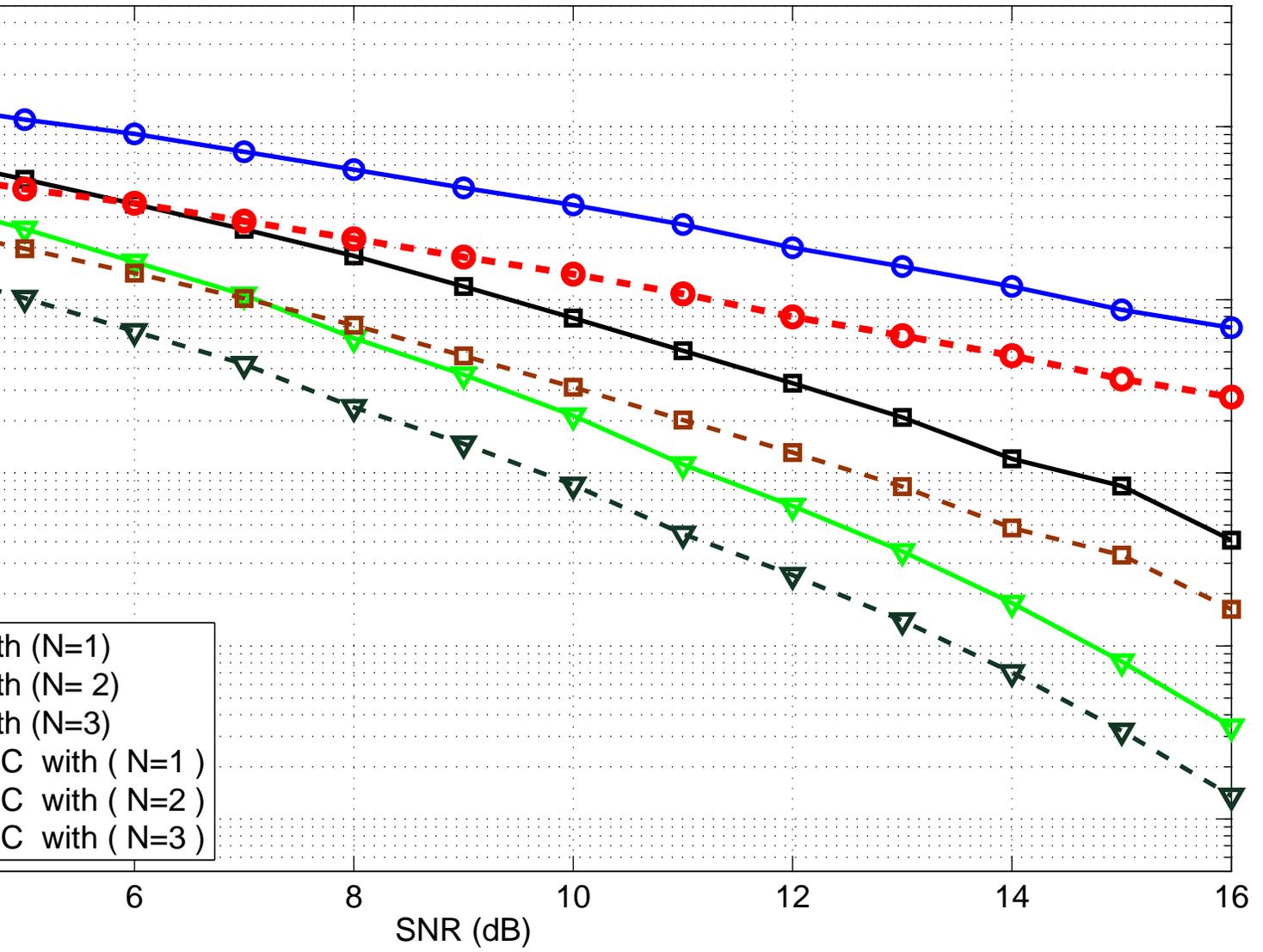

00 , 20 , 02 , 22 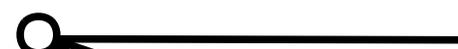  00 , 01 , 02 , 03

01 , 21 , 03 , 23 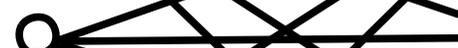  10 , 11 , 12 , 13

11 , 31 , 13 , 33 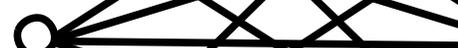  20 , 21 , 22 , 23

12 , 32 , 10 , 30 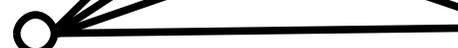  30 , 31 , 32 , 33

(a)          (b)

0000 , 2030 , 0012 , 2022

0101 , 2131 , 0113 , 2123

1201 , 3231 , 1213 , 3223

1230 , 3332 , 1310 , 3320

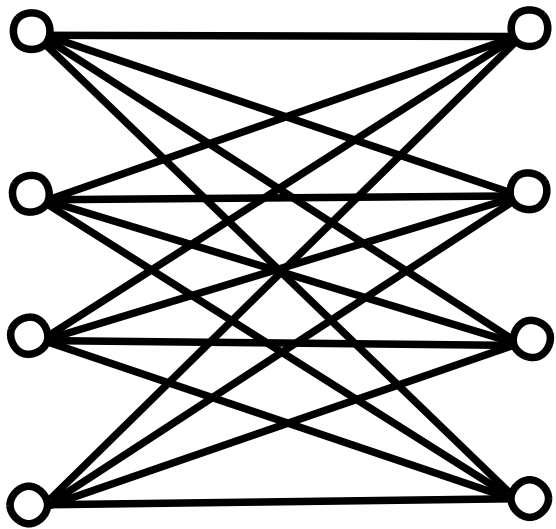

(a)

0000 , 0001 , 0002 , 0003

1011 , 1012 , 1013 , 1010

2021 , 2122 , 2223 , 2320

3032 , 3133 , 3130 , 3231

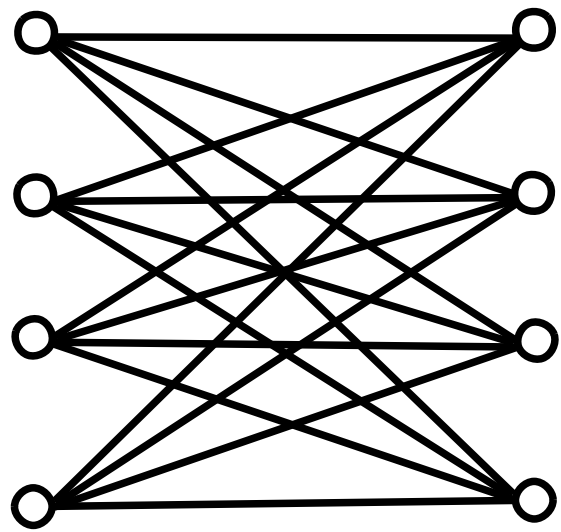

(b)

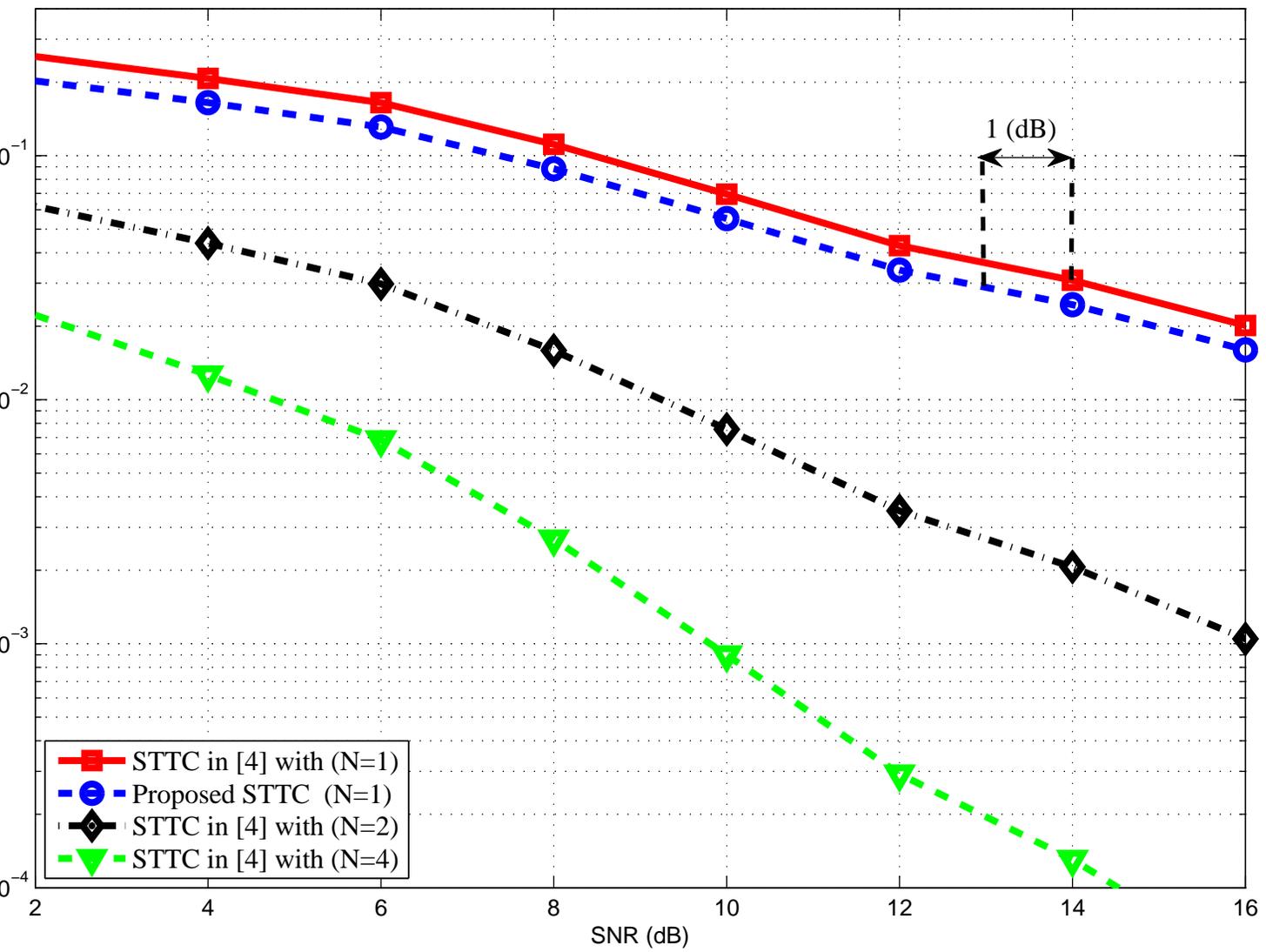